# Ultrathin catalyst-free InAs nanowires on silicon with distinct 1D sub-band transport properties

F. del Giudice[1], J. Becker[1], C. de Rose[1], M. Döblinger[2], D. Ruhstorfer[1], L. Suomenniemi[1], H. Riedl[1], J. J. Finley[1], and G. Koblmüller[1]

[1] *Walter Schottky Institute and Physics Department, Technical University of Munich, Garching, Germany*
[2] *Department of Chemistry, Ludwig-Maximilians-University Munich, Munich, Germany*

Ultrathin InAs nanowires (NW) with one-dimensional (1D) sub-band structure are promising materials for advanced quantum-electronic devices, where dimensions in the sub-30 nm diameter limit together with post-CMOS integration scenarios on Si are much desired. Here, we demonstrate two site-selective synthesis methods that achieve epitaxial, high aspect ratio InAs NWs on Si with ultrathin diameters below 20 nm. The first approach exploits direct vapor-solid growth to tune the NW diameter by interwire spacing, mask opening size and growth time. The second scheme explores a unique reverse-reaction growth by which the sidewalls of InAs NWs are thermally decomposed under controlled arsenic flux and annealing time. Interesting kinetically limited dependencies between interwire spacing and thinning dynamics are found, yielding diameters as low as 12 nm for sparse NW arrays. We clearly verify the 1D sub-band structure in ultrathin NWs by pronounced conductance steps in low-temperature transport measurements using back-gated NW-field effect transistors. Correlated simulations reveal single- and double degenerate conductance steps, which highlight the rotational hexagonal symmetry and reproduce the experimental traces in the diffusive 1D transport limit. Modelling under the realistic back-gate configuration further evidences regimes that lead to asymmetric carrier distribution and lifts of the degeneracy in dependence of gate bias.

## 1. Introduction

Indium arsenide nanowires (InAs NWs) have emerged as a unique and widely studied class of one-dimensional (1D) small band-gap III-V semiconductors owing to several key characteristics, such as small electron effective mass and high electron mobility, wide Bohr radius, strong spin-orbit interaction as well as large Landé factor [1]. The resulting 1D electronic properties provide an important foundation for future devices in nanoelectronics, topological quantum information processing, and energy harvesting applications. For instance, 1D-like InAs NWs are exploited in hybrid semiconductor-superconductor junctions in the quest for Majorana fermions [2,3]. They are also heavily studied as potential candidates for downscaled transistors both in tunneling field effect transistors (TFET) [4] and complementary metal-oxide-semiconductor (CMOS) circuits due to the high carrier mobilities and saturation velocities [5,6]. Given the long mean free path of electrons in 1D-InAs NWs, the scaling efforts towards small size dimensions may further enable future ballistic transistors, where transport occurs in distinct 1D-subbands [7,8]. Herein, well separated 1D subbands with sufficiently large energy splitting are much desired for the observation of clear conductance steps, not limited by thermal and disorder-mediated energy broadening [8-10]. This has clearly propelled efforts in realizing ever thinner NWs into the sub-40 nm diameter regime, as the subband splitting increases inversely with diameter [9]. In this regard, the sub-band population was probed through experiment and simulation in idealized cylindrical InAs NWs with diameters as low as ~15-35 nm [9], whereas ballistic transport was examined in both cylindrical and rectangular NWs with cross-sections of ~30 nm over channel lengths up to ~300 nm [7,8]. Moreover, strongly 1D-quantum confined InAs NWs with such small diameters have shown unique potential in nanothermoelectrics, where the thermopower can be modulated through the peaked 1D density of states [11].

To date, the realization of ultrathin (~sub-30 nm) InAs NWs was primarily achieved by vapor-liquid-solid (VLS) growth processes using foreign catalysts, amongst which gold (Au) [11-17] and other metals such as Ag [18] and Ni [19] have been used. A common problem encountered in direct growth from metal catalysts is the typically large diameter and length distribution [13-19] and non-epitaxial growth [13,18,19]. In addition, Au catalysts, which are most heavily used, are incompatible with Si-based CMOS technology due to deep level traps they form in the band gap of Si [20]. Alternative methods that meet CMOS-compatible criteria while simultaneously inhibiting size dispersion effects aim at catalyst-free and selective area epitaxial (SAE) growth, allowing monolithic and epitaxial integration on Si platform. Although a vast amount of SAE-type growth studies of catalyst-free InAs NWs exists [21-25], very little focus was placed on the creation of ultrathin NWs so far. Hereby, a major challenge mediated by the underlying non-VLS type SAE growth is to overcome the lateral growth on the NW sidewall surfaces that naturally leads to increased NW diameters. More recently, template-assisted selective epitaxy (TASE) has been introduced as a unique scheme [4,8,26,27] for III-V NWs on Si and SOI (silicon-on-insulator) where the lateral extension of NWs is restricted by 'forced' growth into confined $SiO_2$ nanotube templates (vertical [26] or in-plane [8,27]). This method allows reproducible fabrication of sub-30 nm thin, oxide-encapsulated InAs NWs with clear 1D-transport characteristics, but at the expense of very elaborate lithography processes.

In this work, we develop direct catalyst-free SAE growth of ultrathin InAs NWs on Si where optimized growth parameters and intimate dependencies between NW spacing, mask opening and growth time are exploited to realize NW diameters close to ~20 nm. Based on such high-uniformity catalyst-free (111)-oriented InAs NWs we further propose an inverse growth method by which even thinner NWs, as low as ~12 nm, can be created under controlled thermal decomposition of the hexagonal {1-10} NW sidewall surfaces. To demonstrate the presence of discrete 1D-subbands, we finally characterize pronounced conductance steps in low-temperature gate voltage dependent transport measurements which are further supported by simulations of the sub-band population in the realistic hexagonal NW geometry. Moreover, we show how the charge carrier distribution in the individual 1D-subbands evolves with gate voltage, identifying regimes where the degeneracy of states even breaks down due to asymmetries in the carrier distribution.

## 2. Results and Discussion

### 2.1. Synthesis of ultrathin InAs nanowires

In our studies, we used molecular beam epitaxy (MBE) to synthesize vertically oriented InAs NWs on Si (111) via catalyst-free and position-controlled SAE growth processes. As further described in detail in the Experimental Section, the SAE-type growth was facilitated by pre-patterning $SiO_2$-templated Si(111) substrates with regular arrays of mask openings of specific diameters and separation (pitch) using electron beam lithography (EBL). Based on such templated growth, in the following we demonstrate two different growth procedures to realize ultrathin, epitaxial InAs NWs with diameters tuned to the extreme size limits; i.e., (i) direct bottom-up vapor-solid (VS) growth under growth parameters optimized for high aspect ratio NW arrays, and (ii) reverse-reaction growth (RRG) on high-uniformity as-grown NW arrays where the post-growth annealing time under continuous As supply is tuned to accurately control the NW thinning processes.

2.1.1. Bottom-up MBE growth of ultrathin InAs nanowires

We first present investigations towards ultrathin InAs NWs via direct catalyst-free selective area epitaxial (SAE) growth. Previous efforts in the catalyst-free growth of InAs NWs have shown that forming NWs with high aspect ratio (at minimum radial growth) depends sensitively on growth parameters, such as temperature, the incoming fluxes, and V/III ratio [21,22,29-32]. In particular, high V/III ratio and high growth temperature (below the onset of In desorption) are favored for high aspect ratio NWs, however, irrespective of the employed growth method NW diameters below ~40 nm were hardly explored [21,22,29-32]. Many optimization efforts were further hampered by very large length / diameter and NW density fluctuations, because the majority of growth studies were performed in a self-assembled manner on unpatterned substrates [21,29-33]. Here, we adapt SAE growth of highly periodic NW arrays under optimized growth parameters and further introduce two essential factors, i.e., NW-spacing and growth time, to tune the NWs into the ultrathin (sub-30 nm) diameter regime. The growth was performed in a nominally In-limited growth regime under conditions leading to the highest axial/lowest radial growth rates previously reported for solid-source MBE [33], i.e., growth temperature T = 520°C, an As beam equivalent pressure (BEP) of $4.5\times10^{-5}$ mbar (equivalent to a flux of 24.4 Å/s [31]) and an In flux of 0.6 Å/s (V/III ratio = 40.7).

Figure 1(a) shows an exemplary SEM image of a high-uniformity InAs NW array as obtained after a growth time ($t_G$) of 10 min in mask openings with diameter of 140 nm and a pitch (NW spacing) of 1 µm. The array exhibits an excellent selectivity and yield of > 80% and further shows that the NWs nucleate at the very edge of the openings. This observation is consistent with recent findings of SAE-growth of InAs NWs on $SiO_2$-masked Si (111) by MBE [24] as well as by MOCVD [25]. The NWs shown here are fully untapered and exhibit a length of ~1 µm and a

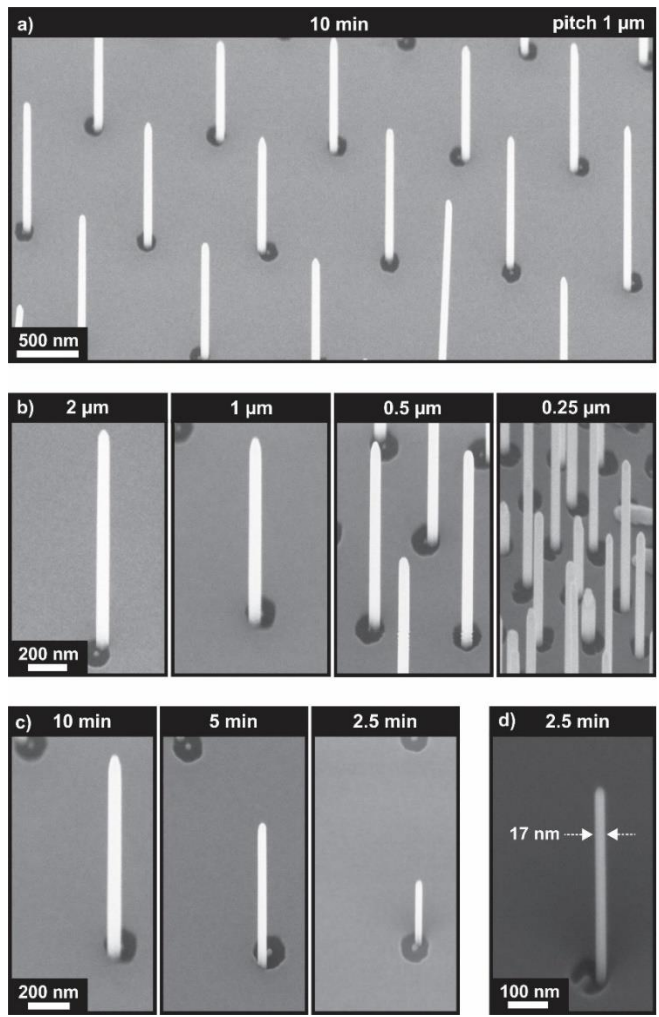

**Figure 1:** SEM images of directly bottom-up grown InAs NW arrays. (a) overview of a typical high-uniformity array after 10 min of growth on a field with 1 µm pitch. (b) representative NWs obtained under the same growth conditions and growth time (10 min) for different pitch ranging from $p = 2$ µm to 0.25 µm (scale bar of 200 nm is the same for all images). In (c) exemplary NWs are depicted for variable growth times of $t_G = 10$ min, 5 min and 2.5 min, obtained from identical fields with 1 µm pitch. (d) ultrathin NW (diameter of 17 nm) when grown from smaller mask opening (~80 nm) for 2.5 min.

diameter of ~60 nm. Slightly reduced diameters of ~50 nm are found for smaller mask openings (see Supporting Information S1), however, at the expense of yield, in agreement with recent studies of SAE-grown InAs [22,25] and InAsSb NWs [34]. Fig. 1(b) illustrates for the same growth time and mask opening diameter (140 nm) how the NW diameter can be tuned by the pitch. In particular, reducing the pitch from $p = 2$ µm to 0.15 µm leads to a decrease in NW diameter from ~65 nm to

~35 nm (see also Fig. 2(b)). Simultaneously, the length of the NWs is reduced from ~1.2 µm to ~0.7 µm, leading to an overall constant aspect ratio (~15-18) over the investigated range of pitch. The pitch dependent behavior points to synergistic effects of surface diffusion on the $SiO_2$ mask and capture length scales of adatoms, governing NW growth. Essentially, when the surface diffusion length of In adatoms on the $SiO_2$ mask is on the order of or larger than half the spacing in between NWs, the NWs tend to compete for adatoms which we refer to as a competitive rate-limiting regime [24,28,35]. As a result, any subtle increase in pitch leads to an increase in the capture area for diffusing In adatoms at the NW and, thus, a larger volume of the corresponding NW. On the other hand, when the pitch increases further such that half the NW-spacing exceeds the surface diffusion length, the competition between neighboring NWs ceases and the NWs become largely independent of each other. In this diffusion-limited regime [24,28,35] hardly any further increase in NW diameter or length is expected with increasing pitch. Such transition to a diffusion-limited growth is, indeed, directly observed in Fig. 2(b), since the NW diameter saturates at a pitch in between ~1-2 µm. This suggests that the surface diffusion length of the migrating (In) adatoms on the underlying $SiO_2$ is in between ~0.5-1 µm, in line with previous studies [24].

Based on these findings, we further tuned the growth time $t_G$ as a critical parameter influencing the NW diameter. Hereby, we performed a series of different growth times ranging from 2.5-min short growths to 60-min long growths under otherwise fixed conditions. Fig. 1(c) depicts SEM images of growths (pitch = 1 µm) from the most interesting short end of this series (10-min, 5-min, 2.5-min), where the thinnest NWs are expected. For the shortest growth of $t_G$ =2.5 min the diameter becomes as small as ~25 nm at a length of about 400 nm (aspect ratio of ~16). To better understand the growth dynamics of these thin InAs NWs and put them into relation with existing literature data of catalyst-free VS-grown InAs NWs on Si, we plot in Fig. 2(a) the time-dependent NW diameter and length evolution obtained from quantitative SEM analysis. The temporal evolution clearly demonstrates strongly non-linear behavior, characterized by an initially rapid increase in NW diameter (radial growth along the <1-10> lateral direction). Likewise, the length of the NW increases continually (axial growth along the <111> direction). Growth progresses then to a highly linear growth regime, where the initial lateral growth is slowed down and where both the NW diameter and length scale linearly for increased growth time beyond 30 min. In the linear regime, the aspect ratio is much higher (up to a factor of 3) compared to growth during the initial phase due to the much lower radial growth rate (see inset). We further note that very similar values of the aspect ratio within the two distinctly different growth regimes were found for all other growths on fields with different pitches ($p = 0.15 – 2$µm).

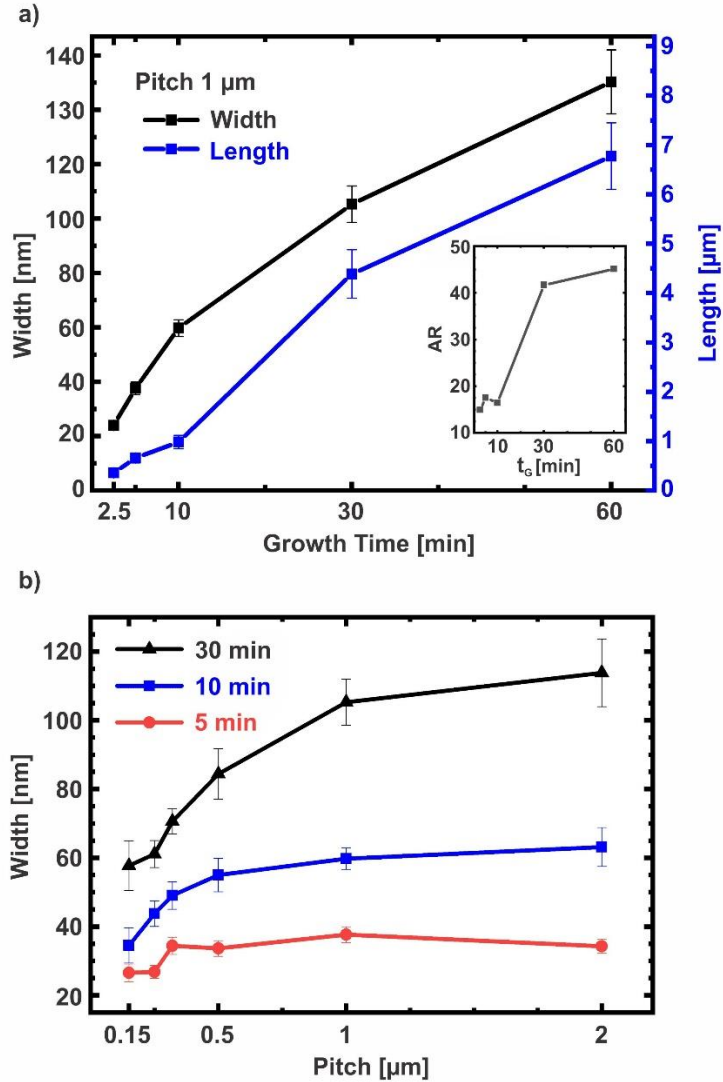

**Figure 2:** (a) Evolution of NW diameter and length as a function of growth time $t_G$ for growth on a pattern with $p = 1$ µm and mask opening of 140 nm. The inset shows the time evolution of the corresponding aspect ratio. (b) Dependence of NW diameter on pitch for three growth times of $t_G$ = 5 min, 10 min and 30 min. For the respective NW length dependenc we refer to the SI section. Error bars stem from the statistical size variation determined by SEM from >10 NW/growth field.

In analogy with previous studies [12,25,36,37], we believe that the presence of the two regimes are governed by differences in the adatom collection and surface diffusion processes. At the beginning of growth, where the NW is relatively small, the collection of In adatoms migrating from the SiO$_2$ mask is very high, leading to strongly non-linear growth behavior. For increased growth time, however, the lateral extension of the NW reduces the collection of adatoms via the mask and, instead, adatom diffusion on the sidewall surfaces become more dominant, resulting in steady-state linear growth. As such, one can also understand the initial growth phase leaning more towards In-rich conditions which are known to yield lower aspect ratio, whereas reduced In collection (present for extended growth time) may lean towards As-rich conditions enhancing axial growth [32]. Finally, our goal was to increase the aspect ratio during the initial growth phase. Motivated by the mask opening dependent data of Fig. S1 (Supporting Information), we therefore combine short growth with reduced mask opening diameter to produce even higher aspect ratio. Using a mask opening diameter of ~80 nm for a growth time of $t_G$ = 2.5 min resulted in ~17-nm thin and ~400 nm long NWs (Fig. 2(d)), yielding a high aspect ratio of ~24. This is by far the lowest ever reported diameter at reasonably high aspect ratio in catalyst-free InAs NWs. Similar studies performed during the initial growth phase of VS-grown InAs NWs showed that at comparable NW length the diameters were at least > 80 nm wide (i.e., aspect ratio < 6) [24,25,30].

2.1.2 Reverse-reaction growth of ultrathin InAs nanowires

In the following, we present another alternative method to create ultrathin (sub-20 nm) InAs NWs that are epitaxially oriented on the Si (111) substrate and which provide lengths exceeding several µm. Specifically, we explore a unique *reverse-reaction growth* (RRG) scheme which constitutes an inverse growth mechanism facilitated by post-growth thermal decomposition of as-grown NWs [38]. This growth mechanism was recently pioneered by us for epitaxial GaAs NWs under in situ annealing experiments under standard MBE ultra-high vacuum (UHV) conditions, producing NWs with diameters below 10 nm [38]. To apply the RRG scheme to InAs NWs we use very high uniformity NW arrays and systematically study the competing thermal decomposition behavior from both the {1-10} sidewall surfaces and the <111> oriented growth front in dependence of array pitch, annealing time and As$_4$ overpressure. As a starting base for these experiments, InAs NW arrays were grown for 30 min at 520 ºC under the same optimized growth parameters as in the previous section. In addition, we introduced a specific pretreatment of the Si (111) substrate surface prior to growth to stabilize an As-terminated Si (111) surface [39] and guarantee NWs arrays with consistently very high yield (>95%) for all investigated pitches

(0.25 – 2µm) (see Supporting Information, Fig. S2). Although this procedure has a profound impact on NW yield, it overall produces shorter NWs as compared to growth without the pretreatment. As expected, the resulting NW lengths / diameters depend sensitively on pitch, ranging from ~1.1-µm-long / 70-nm-wide NWs at $p = 0.25$ µm (A.R. ~15) to ~ 2.7-µm-long / 120-nm-wide NWs at $p = 2$ µm (A.R. ~23) (see Fig. S2). Figure 3(a) shows a characteristic SEM image (left) of the as-grown InAs NW reference for $p = 0.5$ µm, illustrating an average NW length of ~1.8 µm and diameter of ~100 nm. Based on such as-grown InAs NW reference samples, we subsequently performed post-growth annealing directly in the MBE growth chamber by rapidly ramping the growth temperature right after growth termination from 520 °C to 610 °C under the same $As_4$ BEP used for growth. Once the annealing set temperature was reached (within 2 min), the $As_4$ BEP was lowered to $1\times10^{-5}$ mbar to perform annealing under different annealing times. We have selected the final temperature (610 °C) and As overpressure carefully, based on recently identified thermal stability criteria for [111]B-oriented InAs NWs on Si [33]. Herein, it was recognized that InAs NWs become thermally unstable at temperatures above ~560 °C when annealing under pure UHV conditions (base pressure < $10^{-10}$ mbar, without $As_4$ BEP). As a result, NWs decompose rapidly via excessive As desorption leaving behind large metallic In droplets on the growth surface due to the much lower equilibrium vapor pressure of In [33]. To avoid such incongruent desorption behavior and formation of non-stoichiometric surfaces, we employ in the present study a well-controlled As overpressure which slows down As desorption, and, thus, high annealing temperature in excess of 600 °C can be applied.

Fig. 3(a) shows SEM images from NW fields with pitch $p = 0.5$ µm resulting after three different annealing times ($t_{anneal}$) of 10 min, 20 min, and 30 min, respectively. For such small NW spacing, we find that instead of the intended NW thinning, the diameters remain largely unaffected whereas the NW length decreases substantially with annealing time; more than half of the original NW length is lost at 30 min. Concurrently, we observe a shape transformation from untapered NWs to characteristic pencil-shaped NWs upon annealing. This observation hints to the simultaneous thermal decomposition from both the top surface and sidewall surfaces, as discussed below. Some NWs also exhibit formation of very thin tips (< 30 nm) with lengths varying from few tens of nm up to ~300 nm. Closer view by SEM and TEM (Fig. 4) shows that no metallic In, e.g. in the form of droplets, has formed on the NW or the underlying substrate, in contrast to annealing experiments without As overpressure [33] (see also Supporting Information, Fig. S3). In Fig. 3(b) we further illustrate the NW morphology evolution and corresponding RRG characteristics as a function of pitch for a fixed $t_{anneal}$ = 20 min. Most strikingly, we find that while the formation of thick pencil-like NWs is aggravated towards smaller pitch (0.25 µm), larger pitch of $p$ =1 µm and 2 µm results in NWs with diameters that continuously decrease with pitch. Although the tapering still persists in NW arrays with large pitch, the thinned region emerging from the tip is substantially extended in length, i.e., up to several hundreds of nm to > 1 µm. This trend continues further with increasing pitch and annealing time, illustrated in Fig. 3(c) for $p$ = 3 µm and $t_{anneal}$ of 25 min. The resulting NW has a total length of ~2.2 µm and an exceptionally narrow diameter over almost the entire length. More the 50% of the NW has a diameter below 30 nm, which is progressively narrowed to a width as small as ~12 nm in the upper region of the NW.

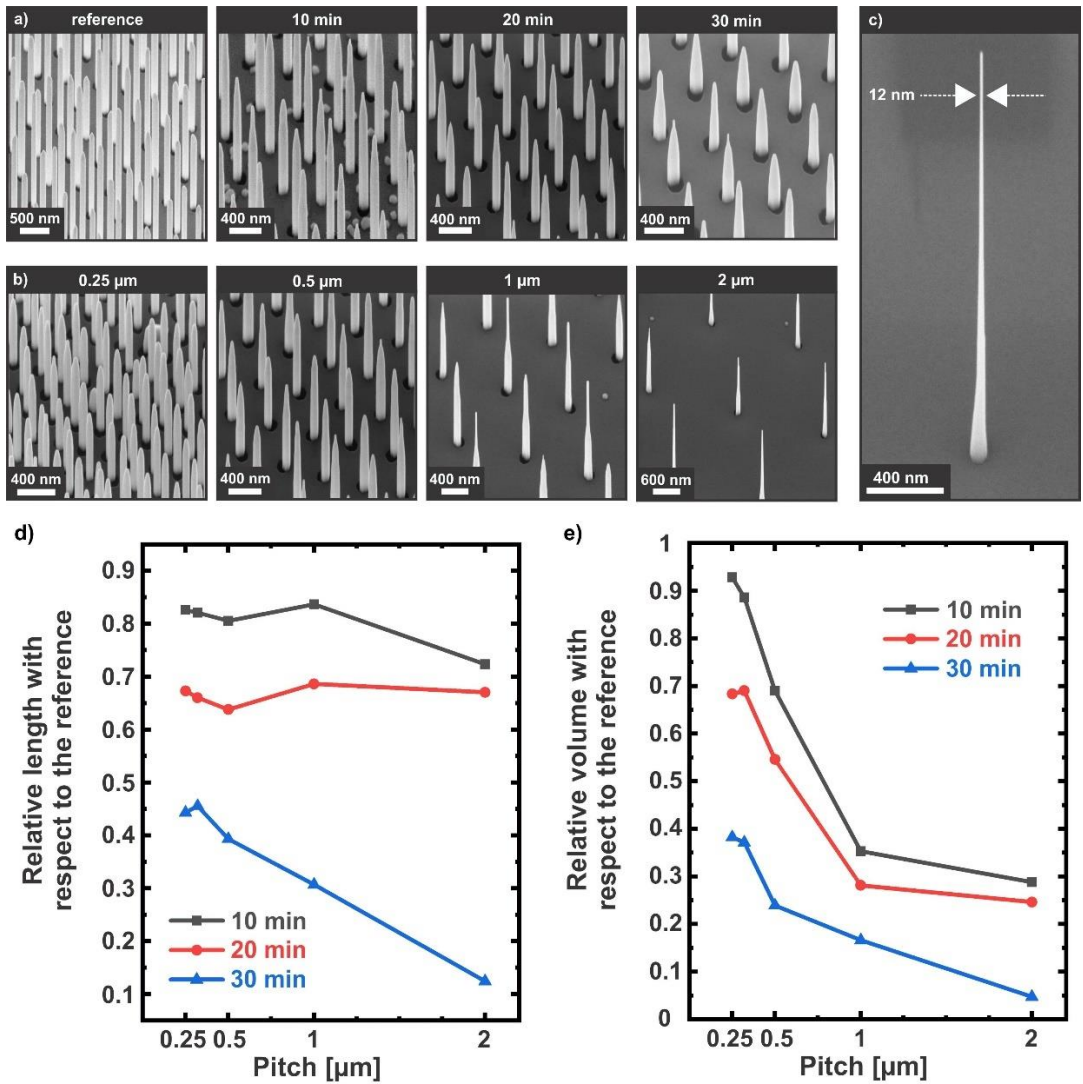

**Figure 3:** (a) SEM images of InAs NW arrays with pitch of $p$ = 0.5 µm before in situ annealing (left) and after applying different annealing times (10 min, 20 min, 30 min). (b) Dependence of pitch on the NW morphology for a constant annealing time of 20 min. (c) Exemplary NW from an array with pitch of 3 µm after annealing time of 25 min. (d) and (e) show the relative change of NW length and volume with respect to the unannealed NW reference sample.

Figs. 3(d) and (e) depict a quantitative summary of the NW morphology evolution with increasing annealing time by plotting the relative changes in NW length and volume with respect to the unannealed NW reference as a function of pitch. We note that the NW volume, instead of the diameter, is a better metric to describe the thermal decomposition dynamics due the underlying tapering. Herein, the NW volume was estimated by measuring the widths at different incremental lengths for a large number of individual NWs and approximating their shape as truncated cones. The data clearly shows the very different axial *vs.* radial decomposition dynamics of the NWs in dependence of pitch. For annealing times up to 20 min we see that the NW length decreases steadily with time, resulting in a NW length reduction by about 30-35% with respect to its original length. The length reduction is, however, nearly independent of pitch for $t_{anneal} < 20$ min. In contrast, for the same range of annealing time the corresponding loss in NW volume is very strongly depending on pitch. For small pitch (e.g., $p = 0.25$-$0.32$ µm) the relative reduction in NW volume is nearly identical to the relative decrease in NW length. This evidences that the change in volume is dominated by thermal decomposition along the axial direction. For larger pitch ($p > 0.5$ µm) the losses in NW volume are much more substantial (up to 70% of the original NW volume at $t_{anneal} < 20$ min) compared to the change in NW length. This clearly demonstrates that the drastic volumetric reduction is mediated by a significant loss in NW diameter, i.e., thermal decomposition in the radial direction. Slight deviations from this behavior are observed for the longest annealing time ($t_{anneal} = 30$ min), where the NW length is also continually decreased at increasing pitch. We suggest this observation arises from the complex interplay between the tapered NW morphology and the excessive radial decomposition in the limit of very long annealing time. It is very likely that the major (thin top) region of the NW has vanished by complete radial shrinkage, leaving behind only the remaining thicker bottom part of the slightly tapered NW. This assumption is corroborated by the experiment shown in Fig. 3(c), where slightly reduced annealing time (25 min) maintained still very long NWs, but at diameters reaching the extreme, i.e., ultra-narrow size limit.

To further explain the observed NW length and diameter evolution upon annealing, we need to consider both thermodynamic and kinetic effects in the thermal decomposition processes. Thermodynamically, the different facets constituting the hexagonal NW crystal are well known to underlie a distinct hierarchy in terms of thermal stability. In particular, in the polar III-V semiconductors the (111)B facet is the thermodynamically least stable facet followed by the (110) facet, whereas the (111)A facet is the most stable facet [40]. The favored growth direction of our InAs NWs, or III-V NWs in general, is the [111]B orientation (i.e., [000-1] in the equivalent wurtzite notation) with sidewall surfaces terminated by the {1-10} family of facets [41-43]. This is directly confirmed by the high-resolution scanning transmission electron microscopy (STEM) image below in Fig. 4, evidencing As-polarity along the [111] growth direction. Hence, thermal decomposition along the [111]B growth axis is expected to dominate with faster decomposition rate as compared to the NW sidewall facets. This is clearly seen by the much more rapid reduction in NW length (i.e., by ~30-50 nm/min) as opposed to the shrinkage in NW diameter which proceeds at rates by at least ~1-2 orders of lower magnitude. Regarding the response on NW diameter, the thermal decomposition of the radial {1-10} facets is strongly dependent on NW-spacing (pitch), i.e., low but steady rate for large pitch whereas decomposition is completely inhibited in the limit of small pitch (cf. Fig. 3(a)). This obviously suggests that additional kinetic effects play a substantial role in the thermal decomposition. In particular, we believe that adatoms desorbing from the rapidly decomposing axial growth plane constitute a residual source of vapor phase atoms assisting the stabilization of the major {1-10} surfaces. Especially, As adatoms with their much higher equilibrium vapor pressure are expected to provide an additional overpressure in NW arrays with very close spacing, since the amount of As emitted from the NWs is much higher when the NW density is large. Consequently, the excess As helps to suppress thermal decomposition of the overall more stable {1-10} sidewall facets in highly packed NW arrays. Indeed, such pathways for As emission from NW growth surfaces and formation of a secondary source for growth has been proposed recently by Ramdani, et. al [44] for the growth of high-density self-catalyzed GaAs NWs. Under these assumptions it is then intuitive that for very sparse arrays with low-density, isolated NWs the effective As overpressure is substantially reduced, allowing progressive decomposition of the sidewall surfaces (cf. Fig. 3(b)). As a result, not only the NW diameter, but the whole NW volume is drastically reduced for large NW spacing (Fig. 3(e)). This pitch-dependent behavior observed in the reverse-reaction growth is, in unique terms, the *exact inverse* behavior of the forward growth reaction where the NW volume increases with NW spacing (Figs. 1(b) and 2(b)). In both cases the driving mechanisms are kinetically limited, i.e., collection and capture of diffusing In adatoms driving the forward growth dynamics [24,28,35] and the anticipated secondary As source inhibiting the reverse reaction. Verification of and further insights into the microscopic pathways leading to the pitch-dependent inhibition *vs.* enhancement in the radial decomposition strongly motivates for future real-time *in situ* studies and modelling.

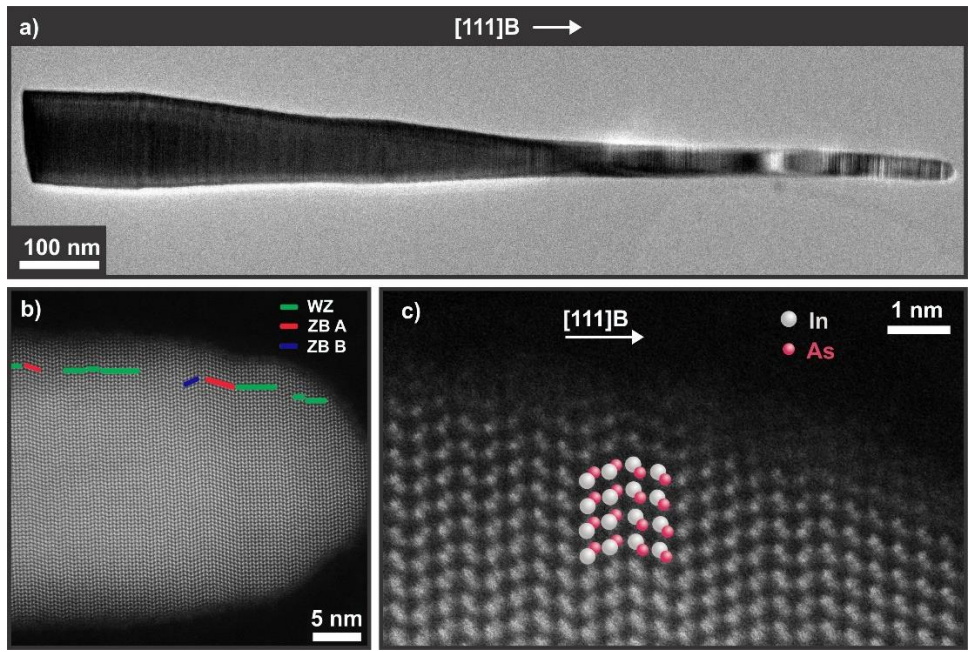

**Figure 4:** Representative TEM micrographs of a NW obtained after the post-growth RRG procedure (annealing time = 20 min, $p$ = 1µm). (a) shows an overview micrograph of the entire NW displaying a large density of stacking defects along the length. The high-resolution micrograph in (b) captures the region near the ~25-nm thin NW tip. Layer stackings associated with WZ domains as well as ZB-type rotational twin domains (ZB-A, ZB-B) are labelled in green, red, and blue, respectively. (c) HR-STEM micrograph from the same region with individual atomic layers depicted in color (In atoms – white, As atoms – red).

Figure 4 shows the typical microstructure of InAs NWs after RRG, measured by high-resolution (HR-) and scanning transmission electron microscopy (S-TEM) in a FEI Titan Themis 300kV-TEM. The depicted NW stems from an array of $p = 1$ µm that was annealed for 20 min. All images were recorded in a <110> zone axis corresponding to the sidewall facets of the NW. Both the large overview micrograph (Fig. 4(a)) and the HR-TEM image recorded near the NW tip (Fig. 4(b)) illustrate a substantial stacking disorder along the entire length of the NW. The predominant layer stacking exhibits WZ-phase with short segment lengths of < 5nm, interrupted by large density of stacking defects and occasional, very short ZB rotational twin domains. Such type of microstructure mimics exactly the structure commonly observed in as-grown catalyst-free InAs NWs when grown at high growth temperature (i.e., >480 ºC) [43,45]. This means that the microstructure seen in these NWs is set by the original as-grown NWs, and the RRG during annealing induces no further changes. Such insensitivity of the microstructure to annealing is similar to recent observations in the RRG of ultrathin GaAs and GaN NWs [38,46]. Additional HR-STEM measurements, depicted in Fig. 4(c), give direct evidence of the [111]B orientation of the NW growth axis. The image clearly resolves rows of atomic bilayers, illustrated by the alternating In/As dumbbells along the growth direction. As expected, due to the higher atomic number (Z), the In atomic columns appear larger and brighter (illustrated also by the color-coded presentation). The arrangement of In atomic columns at the bottom and As atomic columns in the upper layer of each bilayer row confirms the As-polar [111]B growth orientation.

## 2.2. Characterization of 1D-subband transport in ultrathin InAs NW-FET

Finally, we performed low-temperature transport characterization to probe the anticipated 1D-subband nature of the ultrathin InAs NWs. Hereby, back-gated NW-FETs were employed to probe the transfer characteristics as a function temperature, with their fabrication and measurement details being described in the Experimental Section. A typical device is shown in Fig. 5(a) for a NW that stems from the batch presented in Figs. 1/2, hosting a NW with diameter of 25 nm at a channel length (L) of 175 nm. Fig. 5(a) also plots the corresponding transfer characteristics, i.e., $I_{SD}$-$V_{BG}$ behavior measured at a source-drain voltage $V_{SD}$ = 1 mV and gate voltage step size ($\delta V_{BG}$ = 0.5 V) in the temperature range between 200 K and 4K. The original transfer curves recorded for all temperatures overlap substantially in the pinch-off region (~1V < $V_{BG}$ < ~4.5 V), making it difficult to observe distinctly different features in the individual curves. For better clarity, we therefore offset the curves along the $V_{BG}$-axis in consecutive increments of -1.5V with respect to the unshifted data at 4K. At high temperature, relatively smooth depletion of the charge carrier density is observed in the pinch-off region of the $I_{SD}$-$V_{BG}$ data, while lowering the temperature to <50 K results in the appearance of distinct step-like features. These signatures are a first hint for the presence of quantum-confined transport in the NW and the anticipated depopulation of 1D-subbands [7-10, 48, 49]. In particular, the transfer curve taken at 10 K shows very well-defined step/plateau features typical for 1D- transport, with a total of 4-5 steps observed in the investigated gate voltage range and gate voltage spacings $\Delta V_{BG}$ up to several hundred meV between individual steps. Similar number of steps and step spacings were also found in previous reports of InAs NWs with comparable NW diameter, though under distinctly different cross-sectional geometry [8,9]. Lowering the temperature further to 4K clearly intensifies the step-structure due to reduced thermal broadening, however, the step features are superimposed by additional Coulomb blockade-like resonances. Such behavior is typical at very low temperature where random background potential induces quantum dot-like states [49,50] that are weakly coupled to the 1D propagating modes as the thermal energy of the charge carriers is reduced [10,48].

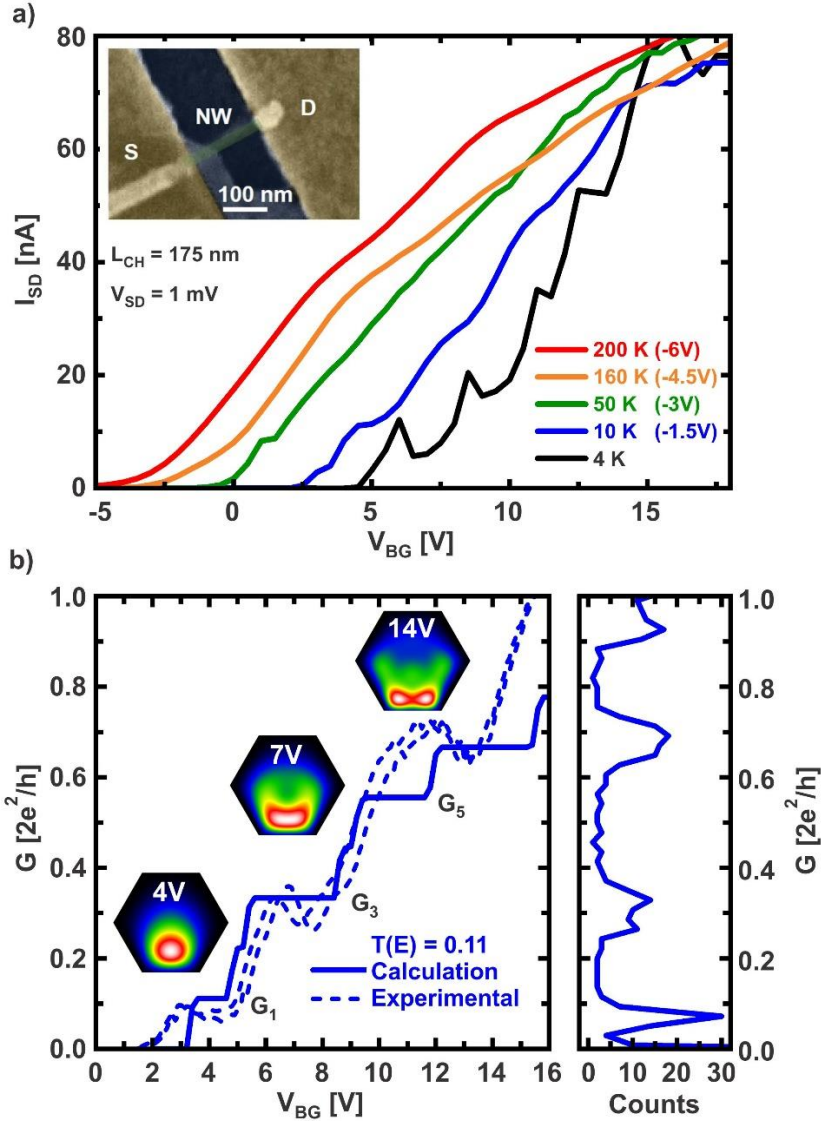

**Figure 5:** (a) Temperature-dependent $I_{SD}$-$V_{BG}$ transfer characteristics recorded at $V_{SD} = 1$ mV for a back-gated NWFET device hosting a 25-nm thin InAs NW at a channel length $L = 175$ nm. The inset shows a color-coded SEM image of the device. For clarity, transfer curves are offset along the $V_{BG}$-axis in -1.5V increments with respect to the data at 4K. (b) Detailed conductance G vs. $V_{BG}$ plot of the same device at 4 K for two $V_{BG}$-sweeps (up- and down-sweep, dashed curves). The step-like features of the experimental data are compared with the simulated conductance G (solid curve) assuming a below-unit transmission probability of $T(E) = 0.11$. The three hexagonal inset graphics illustrate the spatial distribution of charge carriers on a normalized color-scale at three exemplary gate voltages $V_{BG} = 4V$, 7V, and 14V. (Right) Histogram of the observed conductance values obtained from from data binning (50 bins at bin width $\Delta G = 0.02$), showing a series of distinct conductance peaks.

To further elaborate the 1D-transport properties and correlate the distinct step-like features to the underlying subband structure, we plot in Fig. 5(b) in more detail the conductance $G$ as a function of $V_{BG}$ obtained from additional measurements at 4K close to pinch-off. The experimental data plots two gate voltage sweeps (up-sweep/down-sweep) taken at small gate voltage step size $\delta V_{BG} = 0.1$ V). To plots directly illustrate that the step-wise increase in conductance $G$ is fully reproducible and further shows negligible hysteresis. In addition, we correlate the experimental features to the simulated conductance $G$ using the general Landauer formalism for transport in 1D-subbands. Hereby, for the given temperature the conductance is expressed via the current $I = \frac{2e}{h} \int M(E)T(E)(f_S - f_D)\, dE$ in the limit of infinitely small bias voltage, where $M(E)$ is the 1D mode distribution, $T(E)$ the energy dependent transmission function, $f_{S,D}$ the Fermi functions in the S/D ohmic contacts respectively, $e$ the electron charge and $h$ the Planck constant. The Fermi functions take the effects of temperature broadening as well as the influence of the non-zero bias voltage into account, whereas $T(E)$ is assumed to be constant over the energy range of all subbands. In analogy to our previous investigations of 1D-confined GaAs NWs [10,48], we computed the mode density $M(E)$ for each subband spacing by self-consistently solving the Schrödinger-Poisson equations using the Hartree solver nextnano++ [51], shown in the Supporting Information (Fig. S4). Here, we took the realistic hexagonal NW cross-section (width of 25 nm) as well as the experimentally defined back-gate geometry into account. Furthermore, the charge carrier conductivity was assumed to be spatially invariant in the axial direction.

The solid curve in Fig. 5(b) plots the modelled conductance at T = 4 K for the same source-drain bias $V_{SD}$=1 mV used in the experiment. By adapting a transmission probability of $T(E) =$

0.11 we find that the simulated conductance follows closely the traces observed in the experimental data, at least for the lowest 1D states observed near pinch-off. The lower than unity transmission probability is not surprising, given the significant scattering in the NW and the relatively long channel length ($L_{ch}$ = 175 nm) which prevent observation of ideal ballistic transport [7,9,10,48,49,52]. Based on the observed transmission probability and the channel length we estimate the electron mean free path (MFP) $\lambda_e$ to about 76 nm in our 25-nm thin InAs NWs. The MFP is about half the value observed recently in $ZrO_2$-passivated InAs NWs at similar diameter and channel length [7], suggesting that scattering by surface defects in unpassivated NWs is a main source for the comparatively lower transmission probability [53].

Despite the diffusive 1D transport, the simulated $G$ exhibits pronounced steps and plateaus as the Fermi level is tuned through each discrete 1D subband [7-10,48,49], as expected from Landauer quantization. Interestingly, however, unlike in conventional 1D-systems the plateaus do not appear at equidistant steps of $G$ but rather show a sequence of single- and double step heights, i.e., the first plateau arises at $G_1 = 0.11G_0$ ($G_0 = 2e^2/h$), while the second and third plateau emerge at $G_3 = 0.33G_0$ and $G_5 = 0.55G_0$, and the fourth plateau at $G_6 = 0.66G_0$, respectively. Hence, the even multiples of $G_1$ at $2 \cdot G_1$ and $4 \cdot G_1$ are absent, which is ascribed to the rotational hexagonal NW symmetry and the associated two-fold degeneracy of the second and third states [9,10,48,49]. This means that the jumps in steps from $G_1$ to $G_3$ and from $G_3$ to $G_5$ are observed because the density of two subbands have to be depleted due to the degeneracy. The two-fold degeneracy in the second and third step is also reproduced in the experiment, although some slight deviations in $G$ and $\Delta V_{BG}$ spacing occur with respect to the simulated data. This is also evidenced in the histogram (righthand plot of Fig. 5(b)) which depicts the observed conductance values obtained from data binning (50 bins within the conductance range $G = 0$-$1$ at bin width $\Delta G = 0.02$). The conductance peaks at $G_1$ (~$0.1G_0$) and $G_3$ (~$0.3G_0$) are in line with the simulated data, while the third peak comes at a slightly higher $G$ value than the expected. We attribute such variation to the aforementioned superposition by the random background potential (quantum-dot like states) and expected variations in the gate-channel coupling amongst different subbands [48,49]. These are also the reasons why previous reports demonstrated best coincidence between experimental and simulated step/plateau features only for the very first two steps [7,9,49,51].

To directly illustrate the effects of varying gate-channel coupling on the different subbands, we calculated the spatial charge carrier distribution for the first ten eigenstates as a function of the applied $V_{BG}$, under the realistic geometry. For the calculations we assumed infinitely large energy barriers at the NW and an average charge carrier concentration of $6\times10^{18}$ cm$^{-3}$ that is realistic in the limit of ultrathin, unpassivated InAs NWs [54,55]. The entire data illustrating the spatial distribution of the wavefunction amplitudes ($\Psi_n^2$) is of the large number of eigenstates is shown in Fig. S4 (Supporting Information) for a range of backgate voltages $0 < V_{BG} < 14$ V. Exemplary data for the resulting radial charge carrier distribution in the hexagonal NW cross-section is depicted in Fig. 5(b) for $V_{BG} = 4$V, 7V, and 14 V, respectively, i.e., voltages which coincide with the first, second and fourth conductance plateaus. The plots reveal that the charge carriers locate closer to the SiO$_2$ gate dielectric with increasing positive $V_{BG}$. This corresponds to increased electric fields up to e.g., ~$6\times10^7$ V/m in the center of the bottom NW facet and ~$10^8$ V/m at the bottom corners at $V_{BG} = 14$ V, while in the very center of the NW core the respective field is only ~$1.5\times10^7$ V/m. Such highly asymmetric carrier distribution under the influence of a back-gate was recently also seen by Degtyarev, et al. [56], although performed under gate bias of opposite sign. Comparison of the energy eigenvalues (Supporting Information) further shows that states with increasing mode number, i.e., the two-fold degenerate states described by the wavefunction pair amplitudes ($\Psi_2,\Psi_3$), ($\Psi_4,\Psi_5$), ($\Psi_7,\Psi_8$), ($\Psi_9,\Psi_{10}$) have energy eigenvalues relatively close to each other (~17-22 meV) at $V_{BG} = 0$. The energy eigenvalues, however, change to ~ 38 meV for the first two lowest pair amplitudes and ~12 meV and 18 meV for the two remaining pair amplitudes at increased $V_{BG} = 14$V, while the energetic spacing between consecutive pairs is always > 60 meV (see Supporting Information). Particularly, for a given wavefunction pair of a degenerate state one wavefunction is pulled closer to the gate and the other farther away with increasing gate voltage. Comparing these results to simulations of cylindrical NWs (circular cross-section) [9], we can conclude that the hexagonal geometry and especially the asymmetric dielectric environment induced by the back-gate geometry lift the degeneracy of these states by at least 10 meV. This clearly underlines the importance of performing simulations of back-gated NW-FET devices not just at the charge neutrality point but in direct dependence of back-gate voltage.

## 3. Conclusion

In conclusion, we developed ultrathin, monolithic InAs NWs with diameters tuned to the sub-20 nm regime to enable observation of strong 1D quantum confinement effects. Using catalyst-free, site-selective epitaxial growth methods compatible with post-CMOS technology on Si platform, we showed that careful control of growth parameters, interwire separation and mask opening size on $SiO_2$-masked Si (111) facilitates vapor-solid grown NW arrays with diameters as low as ~17-25 nm. In addition, an unconventional reverse reaction growth (RRG) scheme was demonstrated by which post-growth thermal decomposition of as-grown NW arrays under controlled annealing conditions was exploited to realize high aspect ratio InAs NWs with diameters as thin as 12 nm. Hereby, interesting insights into competing thermodynamic and kinetic effects were found based on systematic studies of the array spacing and annealing time. To corroborate the expected 1D confinement behavior, we probed the 1D sub-band structure in ~25-nm thin InAs NWs by low-temperature transport characterization using back-gated NW field effect transistor devices. Clear conductance steps and evidence of single- and double degenerate states were observed, confirming the underlying rotational hexagonal symmetry of the NW. Correlated simulations under the realistic back-gate configuration also highlighted that the charge carrier distribution in the consecutive 1D-subbands is asymmetric and strongly gate voltage dependent, leading to a breakdown of the two-fold degeneracy in higher subbands.

**Experimental Section**

*Synthesis:* All InAs NWs investigated in this study were grown in a solid source Gen-II MBE system with a standard effusion cell for In and a valve cracker cell providing As in the form of uncracked $As_4$. The NWs were grown on commercially available p-type doped Si(111) wafers with a 20-nm thick thermally grown $SiO_2$ mask layer on top. To facilitate site-selective growth the $SiO_2$ layer was prepatterned by electron beam lithography (EBL, Raith eLine) and subsequent reactive ion etching (RIE). Thereby, several periodic mask patterns were created on each wafer which are ordered in hexagonal lattices with circular openings of ~140 nm in diameter (unless otherwise noted) and pitch variable between $p = 0.15 - 2$ μm. Similar to our previous studies [24,28], the patterned substrates were shortly dipped in buffered hydrofluoric acid (HF) to remove residual $SiO_2$ inside the openings and prevent the exposed H-terminated Si (111) surface from re-oxidation.

Subsequently, the substrates were built into the MBE, followed by a prebake at 250°C for at least 2 h and a final cleaning step at 700°C for 20 min right before growth to remove residual surface contaminants. The substrate temperature was measured in situ using an optical pyrometer. Growth proceeded by ramping the substrate to the desired growth temperature under As supply before opening the In shutter. The morphologies and geometrical dimensions of all corresponding NW arrays were characterized by scanning electron microscopy (SEM, Zeiss NVision) under a 45° viewing angle.

*Electrical characterization:* For electrical measurements of the 1D-subband transport characteristics, back-gated NWFETs were fabricated by transferring thin InAs NWs onto ~215-nm thermally grown $SiO_2$ on $n^{++}$-Si that serves as global back-gate. The substrate also contained large Ti/Au bond pads defined by optical lithography. Prior to contacting of individual NWs, the native oxide was removed from the NW surface using an ammonium fluoride/2%-buffered hydrofluoric acid mixed with deionized water (ratio 1:2). Subsequently, standard electron beam lithography, metallization, and lift-off were employed to establish two contact fingers for source/drain (S/D) electrodes (20 nm Ni/80 nm Au) on each NW without further annealing [43,47]. The device was further mounted on a custom-made chip carrier by wire bonding, and then placed into a He-flow cryostat to facilitate temperature-dependent measurements from 300 K to 4.2 K. The back-gate voltage ($V_{BG}$) was applied using a HP 4142B modular DC source, and the resulting source-drain current ($I_{SD}$) data was taken by a EG&G Model 7265 lock-in amplifier at a frequency of 37 Hz.

## Acknowledgements

This work was supported by the Deutsche Forschungsgemeinschaft (DFG, German Research Foundation) under Germany's Excellence Strategy via e-conversion, EXC 2089/1-390776260 and the Munich Center for Quantum Science and Technology (MCQST), EXC 2111-390814868. The authors further acknowledge financial support by the DFG-projects KO4005/5-1, KO4005/6-1, the TUM International Graduate School of Science and Engineering (IGSSE), and the TUM Institute for Advanced Study (IAS) via the Focal Period 2019 program. We further thank the cluster of excellence Nanosystems Initiative Munich (NIM).

## Conflict of Interest

The authors declare no conflict of interest.

## Supporting Information

Supporting Information is available online.

**Supporting Information for**

**Ultrathin catalyst-free InAs nanowires on silicon with distinct 1D sub-band transport properties**

F. del Giudice[1], J. Becker[1], C. de Rose[1], M. Döblinger[2], D. Ruhstorfer[1], L. Suomenniemi[1], H. Riedl[1], J. J. Finley[1], and G. Koblmüller[1]

[1] Walter Schottky Institute and Physics Department, Technical University of Munich, Garching, Germany
[2] Department of Chemistry, Ludwig-Maximilians-University Munich, Munich, Germany

___________________________________________________________________________

1. **Influence of SiO$_2$/Si(111) patterning parameters on mask dimensions and NW growth**

To identify optimum process parameters during the substrate pre-patterning using electron beam lithography (EBL), we investigated the influence of exposure dose of the electron beam on resulting key parameters such as mask opening size, NW yield, and NW diameter. Using conventional PMMA EBL resist we thereby explored a large variation in exposure dose ranging from 2 – 400 fC. Fig. S1(a) shows the impact of exposure dose on the mask opening size $d_0$ for different pitches ranging from p = 0.25 µm to 2 µm. Clearly, it is obvious that higher exposure dose leads to larger mask opening size. Up to a dose of ~20 fC the corresponding increase in mask opening diameter was rather independent of the pitch. However, for larger doses the influence by the pitch becomes quite strong, leading to significantly larger mask opening diameter when the pitch is smaller. This is most likely due to increased exposure in the limit of a high density of exposure spots at narrow pitch. Overall, this allowed us to tune the mask opening diameter from below 50 nm to about 200 nm, as measured by scanning electron microscopy (SEM).

Given the large size tunability of the mask openings, we further recognize a substantial impact on the yield of as-grown InAs NW arrays. As shown in Fig. S1(b), higher exposure dose, i.e., larger mask opening diameter, significantly improves the yield of NWs. As expected from the inter-dependence between exposure dose and mask opening size, we therefore notice a shift of the optimum NW yield with exposure dose for the different pitches. For example, for p = 0.32 µm it requires at least an exposure dose of > 75 fC to achieve a maximum yield of >80%. The equivalent dose for such high yield in arrays with p = 1 µm is much larger, i.e., 400 fC. According to Fig. S1(a) these doses for the two exemplary cases correspond to a mask opening diameter of ~ 120-140 nm. In contrast, with decreasing mask opening size the NW yield drops continuously. Specifically, the NW yield drops to below 20% for mask opening diameters below ~50 nm. Finally, the mask opening diameter also impacts the resulting NW diameter as shown in Fig. S1(c) for an array of p = 1 µm grown for 10 min. We clearly observe that the NW diameter increases continuously for larger opening diameters.

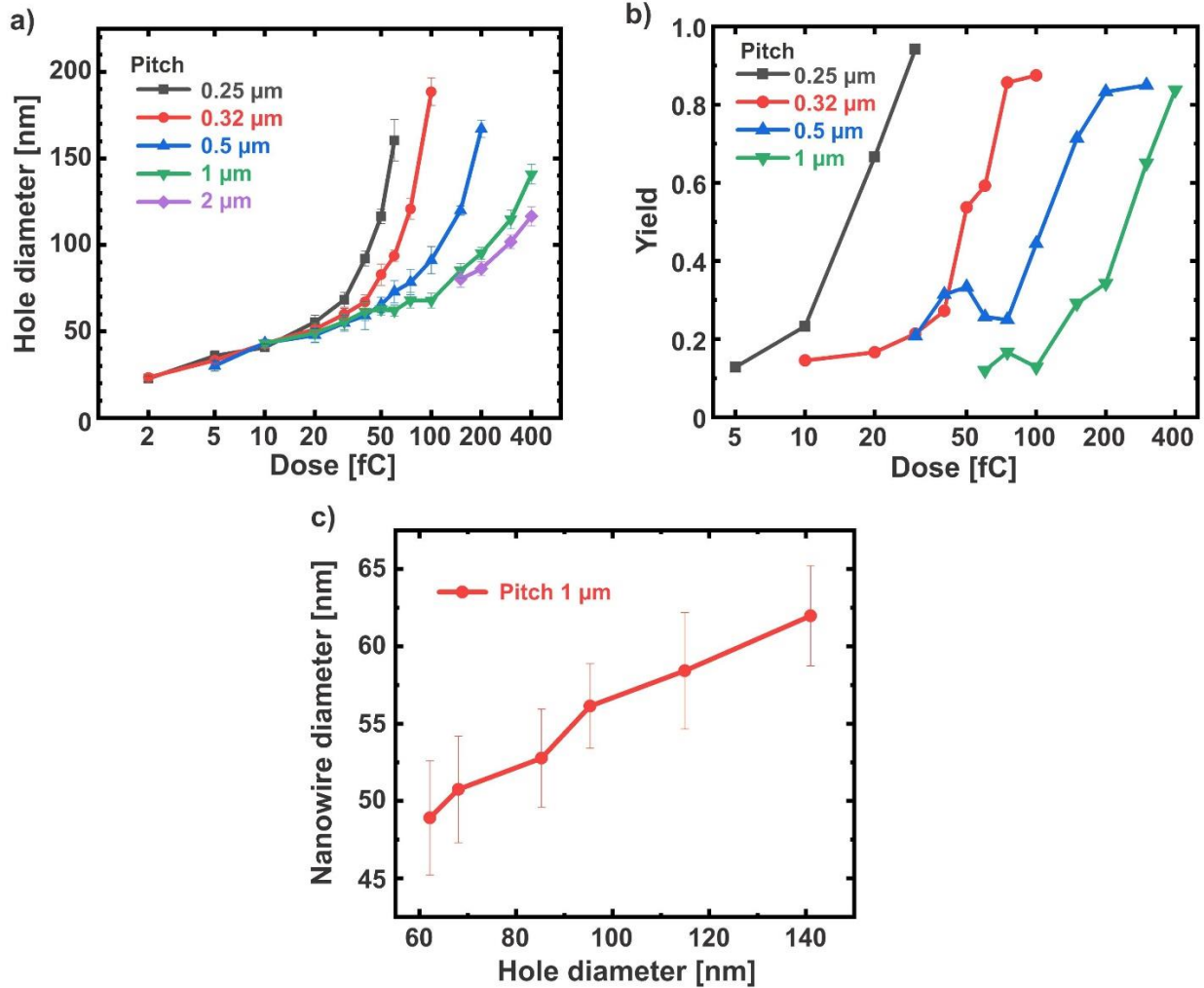

**Figure S1:** Dependence of EBL exposure dose on the mask opening diameter (a) and the resulting NW yield (b) for different pitches. (c) Exemplary evolution of NW diameter as a function of mask opening diameter as obtained for a 10-min long growth of InAs NWs with an array pitch of 1 µm.

## 2. Role of Si(111) surface pre-treatment on NW growth

In order to achieve consistently high yield and high-uniformity in InAs NW arrays we introduced a specific in-situ pretreatment of the patterned $SiO_2$/Si (111) substrate prior to growth. Following a procedure previously established for catalyst-free GaAs NWs grown on Si (111) by selective area epitaxy [1], we apply the following sequences outlined in the process flow diagram of Fig. S2(a): In a first step, the Si substrate was heated to 700°C for 20 min to remove hydrogen (H) atoms from the H-terminated Si (111) surface inside the $SiO_2$ mask openings, which was formed by the HF-dip prior to loading the samples into the MBE. Once the H atoms are removed, the Si (111) surface is expected to modify its surface phase to a (7×7) reconstructed surface. Further heating to 870°C transforms the surface to a (1×1) reconstructed surface phase. To enable high probability of As-polar [111]B-oriented NWs, we supplied an As flux of $4.5\times10^{-5}$ mbar (equivalent to the flux used during growth) to stabilize an As-terminated (1×1)-reconstructed surface. After this, the temperature was ramped to the growth temperature at 520°C and by opening the In shutter InAs NWs were grown for 30 min, forming the base for post-growth annealing experiments as described in the main text. The annealing procedure is also illustrated in the process flow diagram (i.e., temperature ramp to 610 °C and supply of As–flux $1.05\times10^{-5}$ mbar).

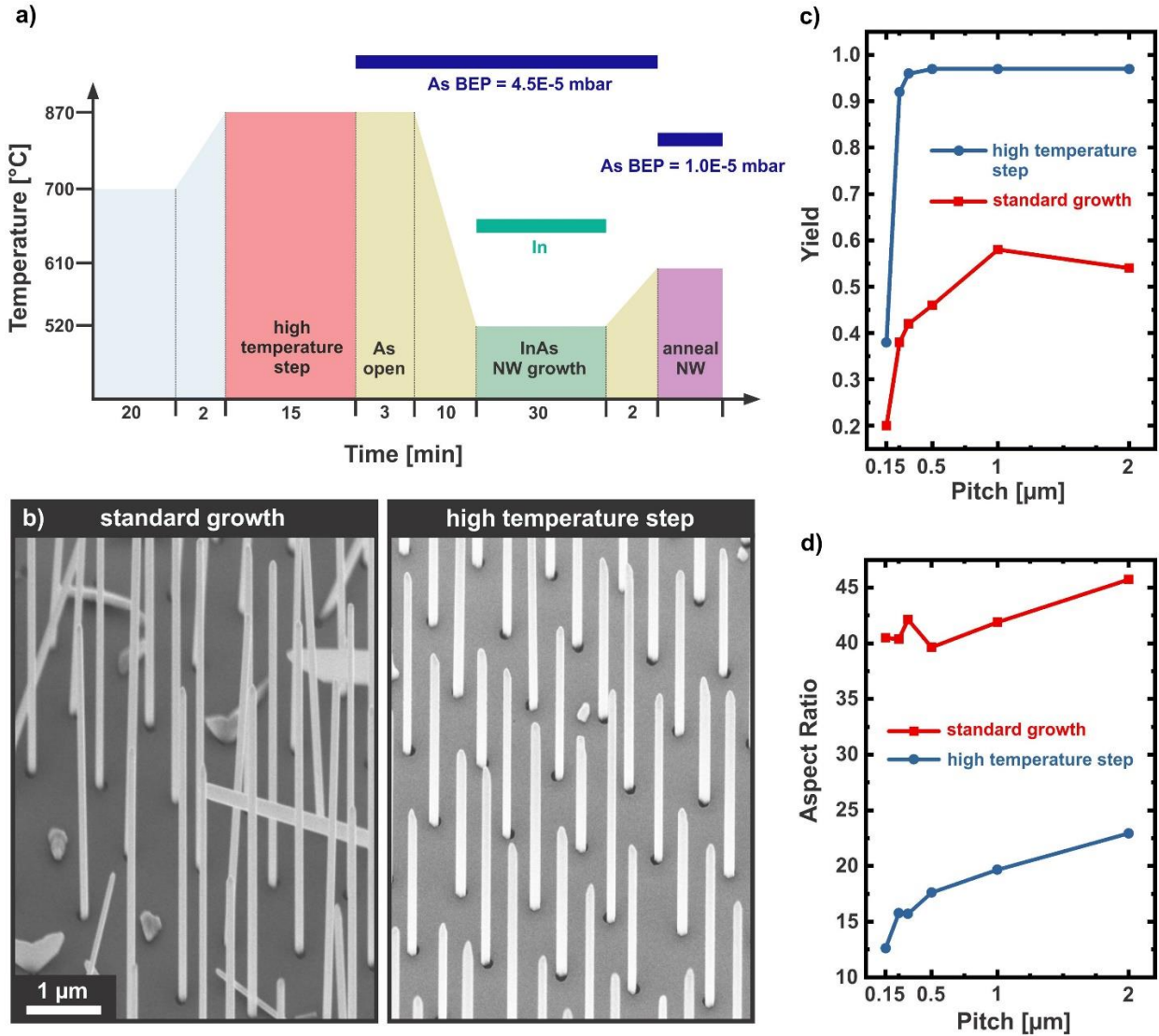

**Figure S2:** (a) Process flow diagram for the in situ pretreatment and successive growth steps of high-yield InAs NWs on $SiO_2$-masked As-terminated Si (111). (b) Comparison of SEM morphologies of InAs NW arrays obtained without (left) and with high-temperature pretreatment (right). The pitch and mask opening diameter for these growths are 1 µm and 140 nm, respectively. NW yield (c) and aspect ratio (d) as a function of pitch for the two growth cases.

Fig. S2(b) compares SEM morphologies of InAs NW arrays right after 30 min of growth on patterned fields with pitch p = 1 µm for the cases with pretreatment (right) and without (left). Clearly, the NW array grown on the pretreated substrate exhibits much higher homogeneity and a yield in excess of 95%. This is further confirmed in the plot of Fig. S2(c), which illustrates the NW yield for different pitches. Consistently, NW yield larger than 90% was observed almost for all investigated pitches, when pretreatment was performed. In contrast, for samples grown without pretreatment the yield was lower (typically less than 80%) along with fluctuations amongst different growth runs. The exemplary sample presented in Fig. S2(c) has a maximum yield of ~60% for p > 1µm, whereas the yield decreases towards lower pitch. Here, we investigated even pitch as low as p=0.15 µm, where the yield drops markedly for both types of samples, and is systematically observed for all growths. Since the yield at such small pitch was too low to unambiguously study the effects of interwire-spacing dependent thermal decomposition dynamics, we limited our investigations in the main text to pitches in excess of 0.25 µm. Furthermore, the pretreatment had an interesting effect on the aspect ratio of the NWs (Fig. S2(d)). While the NW diameters changed only marginally, the NW length was highly affected, leading to much shorter NWs on pretreated Si substrates. Aspect ratios are, hence, about a factor of 2-3 smaller as compared to NW arrays grown on untreated substrates and depend on pitch, in analogy with the observations in the main text.

### 3. Reverse-reaction growth without As overpressure

To illustrate the effect of the UHV environment on the reverse-reaction growth during in situ annealing, we performed a comparative experiment without the use of As overpressure. In this case, after growth the temperature was ramped to 590°C under the same As-flux used during growth ($4.5 \times 10^{-5}$ mbar). Once the set temperature was reached (after 2 min) the As-flux was turned off and different annealing times were applied to explore the reverse-reaction growth dynamics. Fig. S3 on the left shows the typical reference obtained without undergoing the temperature ramp. As expected, the NWs have an original length of several µm with a non-tapered morphology. For 3 min of in situ annealing at 590°C (i.e., without any As-flux) pencil-shaped NW tips start to evolve, similar to the observations made under the presence of As-overpressure in the main text. After another 3 min of annealing, only few NWs remained that became progressively thinner. However, the majority of NWs were decomposed, leaving macroscopic droplet-shaped clusters on the underlying substrate. Additional annealing by another 3 min (total of 9 min) leads to the complete decomposition of all NWs. Only metallic In droplets are visible on the substrate, indicating the non-congruent evaporation conditions under the absence of As overpressure.

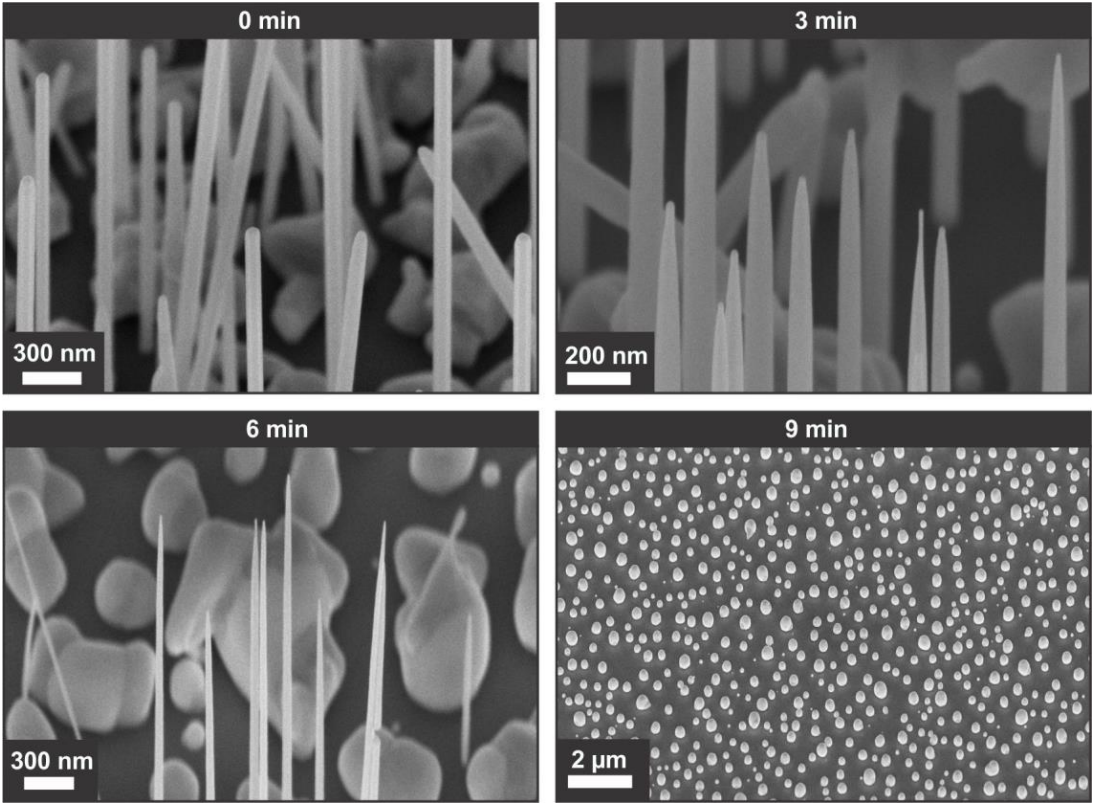

**Figure S3:** SEM morphologies of InAs NWs upon in-situ annealing at 590°C under the absence of As-overpressure. (Upper left) Reference sample prior to annealing. The other images depict morphologies obtained for annealing times of 3 min, 6 min, and 9 min, respectively. After 9 min all NWs are evaporated leaving only metallic In droplets behind.

4. **Energy eigenvalues and radial carrier distribution in dependence of gate voltage**

Using the Hartree solver nextnano++ [2] we self-consistently calculated the spatial distribution of the wavefunction amplitudes ($\Psi^2_n$) for a significant number of eigenstates in dependence of the applied backgate voltage $V_{BG}$. Fig. S4(a) shows the resulting distribution of the squared wavefunction amplitudes in the cross-section of the hexagonal InAs NW (width of 25 nm), where the bottom facet of the NW is anchored to the 215-nm thick $SiO_2$ gate dielectric (not shown in the viewgraphs). In total, these are illustrated for the first 10 eigenstates ($\langle\Psi_n|E|\Psi_n\rangle$) at four different backgate voltages $V_{BG}$ = 0V, 4V, 7V, and 14 V. When looking at the ground state $\Psi_1$ we can clearly see that the center of gravity of the wavefunction amplitude $\Psi_0^2$ is pulled closer to the gate dielectric upon increasing $V_{BG}$.

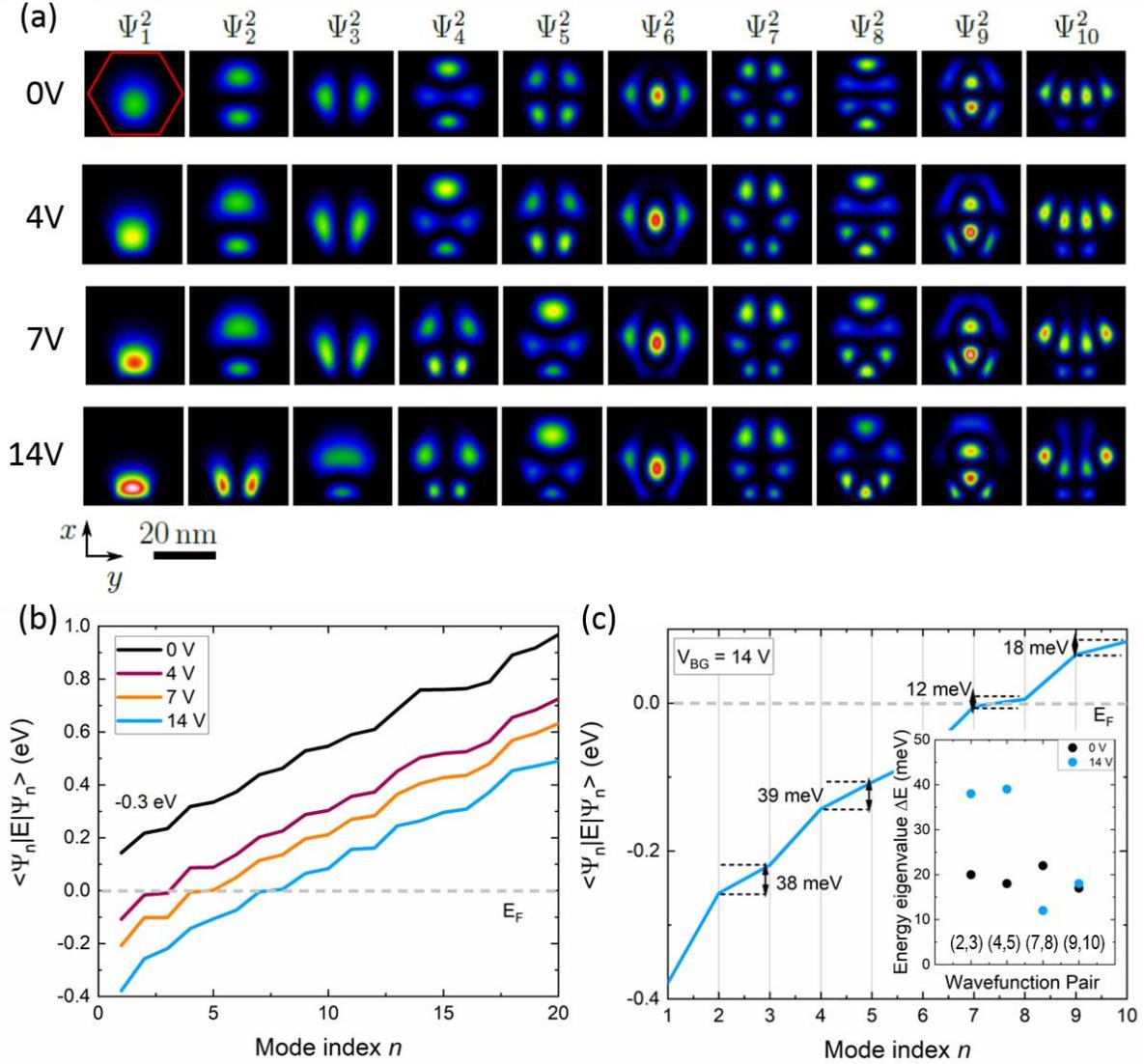

**Figure S4:** (a) Squared electron wavefunction amplitudes $\Psi_n^2(\vec{r}) = \Psi_n^*(\vec{r}) \cdot \Psi_n(\vec{r})$ of the first 10 eigenfunctions in a 25-nm wide InAs NW for applied backgate voltages of 0V, 4V, 7V, and 14V, respectively. The red hexagon in the top left image delineates the cross-section of the NW. The wavefunction amplitudes are ordered according to their energy eigenvalue $\langle\Psi_n|E|\Psi_n\rangle$. (b) Energy eigenvalues versus mode index n for the different backgate voltages $V_{BG}$ = 0V, 4V, 7V and 14 V for the first 20 eigenstates in the same device (the trace for $V_{BG}$ = 0V was shifted by -0.3eV for clarity). The Fermi level $E_F$ is indicated by the dashed line. (c) Detailed view of the energy eigenvalue spectrum for the first 10 eigenstates at $V_{BG}$ = 14V, depicting the different energy separation for the two-fold degenerate states related to the wavefunction pair amplitudes ($\Psi_2,\Psi_3$), ($\Psi_4,\Psi_5$), ($\Psi_7,\Psi_8$), ($\Psi_9,\Psi_{10}$). The inset compares the corresponding energy separation (in meV) for these two-fold degenerate states for gate voltages of 0V (black data) and 14 V (blue data).

For the remaining states we find that pair amplitudes ($\Psi_2,\Psi_3$), ($\Psi_4,\Psi_5$), ($\Psi_7,\Psi_8$), ($\Psi_9,\Psi_{10}$) exhibit energy eigenvalues that are relatively close to each other at $V_{BG} = 0V$, with values of ~20 ± 2 meV for all pairs (mode index 2-3, 4-5, 7-8, 9-10, see Fig. S4(b) and the energy eigenvalue data in inset of Fig. S4(c)). For increased gate voltage these energy values start to deviate, e.g. at $V_{BG} = 14V$ the corresponding values change to ~38 meV for pairs 2-3 and 4-5, while for pairs 7-8 and 9-10 the values are 12 meV and 18 meV (see Figs. S4(c)). As shown in Fig. S4(a), for increased backgate voltage one wavefunction of the pair amplitude is pulled closer to the gate dielectric and the other one further away, increasing the asymmetric carrier distribution. Thus, the degeneracy of the lowest two-fold degenerate states (e.g. 2-3,4-5) at $V_{BG} = 14V$ are lifted almost by up to 20 meV with respect to the case at $V_{BG} = 0$. Also we notice that due to the shifting of the wavefunction towards the gate dielectric with increasing $V_{BG}$, some of the almost degenerate states are swapped. For instance, this is seen for state $\Psi_2(7V) \leftrightarrow \Psi_3(14V)$ and $\Psi_4(4V) \leftrightarrow \Psi_5(7V)$.